\documentclass[10pt]{article}
\usepackage{amssymb}
\usepackage{amsmath}
\usepackage{graphicx}
\usepackage{fancyhdr}
\usepackage{cite}

\begin{document}

\newcommand{\bin}[2]{\left(\begin{array}{c}\!#1\!\\\!#2\!\end{array}\right)}

\huge

\begin{center}
Number of spin-$J$ states and odd-even staggering for identical particles in a single-$j$ shell
\end{center}

\vspace{0.5cm}

\large

\begin{center}
Jean-Christophe Pain\footnote{jean-christophe.pain@cea.fr}
\end{center}

\normalsize

\begin{center}
CEA, DAM, DIF, F-91297 Arpajon, France
\end{center}

\vspace{0.5cm}

\begin{abstract}
In this work, new recursion relations for the number of spin-$J$ states for identical particles in a single-$j$ shell are presented. Such relations are obtained using the generating-function technique, which enables one to exhibit an odd-even staggering in the spin distribution of an even number of fermions in a single-$j$ shell: the number of states with an even value of $J$ is larger than the number of states with an odd value of $J$. An analytical expression of the excess of states with an even value of $J$ is provided, and its asymptotic behavior for large values of $j$ is discussed. 
\end{abstract}

{\bf PACS} 21.60.Cs, 05.30.Fk, 21.30.Fe, 21.10.Hw, 21.60.Fw

\section{\label{sec1} Introduction}

The single-$j$ shell plays an important role in nuclear physics \cite{TALMI93}. Investigations concern in particular symmetries \cite{ZAMICK08}, isospin relations or the $J$-pairing interaction (see for instance Refs. \cite{ZAMICK05d,FU13,ZAMICK07,ZAMICK12,ZAMICK13b}). The single-$j$ shell was also successfully modeled using two-body random Hamiltonians \cite{ZHAO01,MULHALL00}. The enumeration of the number $N(J,j,n)$ of spin-$J$ states for $n$ identical particles in a single-$j$ shell, first adressed by Bethe \cite{BETHE36}, is a fundamental issue of nuclear-structure theory. Such a number can be obtained as \cite{DESHALIT04}:

\begin{equation}\label{rec}
N(J,j,n)=\sum_{M=J}^{J+1}(-1)^{J-M}D\left(M,j,n\right)=D\left(J,j,n\right)-D\left(J+1,j,n\right),
\end{equation}

\noindent where $D(M,j,n)$ represents the distribution of the angular-momentum projection $M$, \emph{i.e} the number of states of a given value of $M$ (I use the notations of Talmi's paper \cite{TALMI05}). There have been many efforts devoted to the determination of an algebraic expression for $N\left(J,j,n\right)$. For instance, Ginocchio and Haxton obtained, in a work on the quantum Hall effect \cite{GINOCCHIO93}, a simple formula for $N\left(0,j,4\right)$, which is also equal to $N\left(j,j,3\right)$. As pointed out by Talmi \cite{TALMI05}, such results are interesting, since it was shown that a necessary and sufficient condition for a two-body interaction to be diagonal in the seniority scheme is to have vanishing matrix elements between $\nu=1$, $J=j$ state ($\nu$ being the seniority) and all $\nu=3$, $J=j$ states of the $j^3$ configuration \cite{DESHALIT04}, and that an equivalent condition is to have vanishing matrix elements between the $\nu=0$, $J=0$ state and all $\nu=4$, $J=0$ states of the $j^4$ configuration \cite{TALMI71}. Zhao and Arima found empirical formulas of $N(J,j,n)$ for three, four and five particles \cite{ZHAO03}. Zamick and Escuderos revisited the Ginocchio-Haxton formula by a combinatorial approach for $J=j$ with $n$=3 \cite{ZAMICK05} and Talmi derived a recursion relation for $N\left(J,j,n\right)$ of $n$ fermions in a $j$ orbit in terms of $n$, $n-1$, $n-2$, \emph{etc.} fermions in a $(j-1)$ orbit \cite{TALMI05}. In Refs. \cite{ZAMICK05b,ZAMICK05c,ZHAO05b}, the studies for $n$=3 and $n$=4 were extended to the number of states with given spin and isospin $T$. In Ref. \cite{ZHANG08}, Talmi's recursion formula \cite{TALMI05} was further generalized to boson systems and applied to prove the empirical formula for $n$=5 bosons given in Ref. \cite{ZHAO03}. The number of states of a given spin was found to be closely related to sum rules of many six-$j$ and nine-$j$ symbols, and coefficients of fractional parentage \cite{ZAMICK05b,ZHAO04,ZHAO05,ZAMICK11,QI10,PAIN11b,WANG12,ZAMICK13,KLESZYK14,HERTZ-KINTISH14}. In Ref. \cite{ZHAO05c}, it was proven that the number of spin-$J$ states for $n$ fermions in a single-$j$ shell or bosons with spin $\ell$ equals the number of states of another ``boson'' system with spin $n/2$, the boson number being equal either to $2j+1-n$ (for $n$ fermions in a $j$ shell) or to $2\ell$ (if one considers $n$ spin-$\ell$ bosons). Jiang \emph{et al.} published analytical formulas for the number of states of a given spin value for three identical particles, in a unified form for both fermions and bosons, by using $n$ virtual bosons with spin 3/2 ($n$ being equal to $2j-2$ for fermions in a single-$j$ shell or to $2\ell$ for bosons with spin $\ell$ \cite{JIANG13}). Recently, Bao \emph{et al.} derived recursive formulas by induction with respect to $n$ and $j$ and applied them to systems of two, three and five identical particles \cite{BAO16}.

In the present work, I propose new recursion relations for $D\left(M,j,n\right)$ obtained using generating functions \cite{SUNKO85,SUNKO86,SUNKO87,KATRIEL89,PRATT00,PAIN11}. The formalism as well as the new relations are described in Sec. \ref{sec2}. The present recurrence relations are different from the one published by Talmi \cite{TALMI05}, but it is shown in the Appendix that the latter can be also easily obtained within our formalism. An approximate statistical modeling of $D\left(M,j,n\right)$ is provided in Sec. \ref{sec3} and compared to exact results. Finally, in Sec. \ref{sec4}, I investigate, still using generating functions, the $J$-excess, which is the difference between the number of states with an even value of $J$ and the number of states with an odd value of $J$. An odd-even staggering for single-$j$ shell with an even number of fermions is observed: the $J$-excess is always positive and is given by a simple binomial coefficient.

\vspace{2cm}

\section{\label{sec2} Generating function and recursion relations}

\subsection{Determination of the generating function}

\noindent Let us consider a system of $n$ identical fermions in a single-$j$ shell (of degeneracy $g=2j+1$) subject to the constraints:

\begin{equation}
n=n_1+\cdots+n_g=\sum_{i=1}^gn_i
\end{equation}

\noindent and

\begin{equation}
M=n_1m_1+\cdots+n_gm_g=\sum_{i=1}^gn_im_i,
\end{equation}

\noindent $m_i$ being the angular momentum projection of state $i$ and $n_i=0$ or 1 $\forall i$. For a configuration $j^n$, one has $J_{\mathrm{min}}=[1-(-1)^n]/4$, 

\begin{equation}\label{jmax}
M_{\mathrm{max}}=J_{\mathrm{max}}=\sum_{m=j-n+1}^jm=\frac{(2j+1-n)n}{2}
\end{equation}

\noindent and $M_{\mathrm{min}}=-M_{\mathrm{max}}$. The corresponding generating function reads

\begin{equation}
f_j(x,z)=\sum_{n=0}^{\infty}\sum_{M=-\infty}^{\infty}z^nx^M\sum_{\{n_1,\cdots,n_g\}}\delta_{n,n_1+\cdots+n_g}~.~\delta_{M,n_1m_1+\cdots+n_gm_g},
\end{equation}

\noindent or

\begin{equation}
f_j(x,z)=\sum_{\{n_1,\cdots,n_g\}}z^{n_1+\cdots+n_g}~.~x^{n_1m_1+\cdots+n_gm_g}.
\end{equation}

\noindent Since the quantities $n_i$ are independent, it is possible to write

\begin{equation}
f_j(x,z)=\sum_{n_1=0}^1z^{n_1}x^{n_1m_1}\cdots\sum_{n_g=0}^1z^{n_g}x^{n_gm_g},
\end{equation}

\noindent which yields

\begin{equation}\label{fjxz}
f_j(x,z)=\left(1+z~x^{m_1}\right)\times\cdots\times\left(1+z~x^{m_g}\right)=\prod_{i=1}^g\left(1+z~x^{m_i}\right).
\end{equation}

\noindent In that case, $D\left(M,j,n\right)$ is related to $f_j(x,z)$ by

\begin{equation}
f_j(x,z)=\sum_{n=0}^{\infty}\sum_{M=-\infty}^{\infty}z^nx^MD(M,j,n),
\end{equation}

\noindent leading to

\begin{equation}
D(M,j,n)=\frac{1}{(2i\pi)^2}\oint\oint\frac{dz_1}{z_1^{n+1}}\frac{dz_2}{z_2^{M+1}}f_j\left(z_1,z_2\right).
\end{equation}

\subsection{New recurrence relations}

The generating function (\ref{fjxz}) can be expanded in powers of $z$:

\begin{equation}\label{eqfj}
f_j(x,z)=\sum_{n=0}^{g}z^nf_{j,n}(x),
\end{equation}

\noindent with

\begin{equation}\label{de}
f_{j,n}(x)=\left.\frac{1}{n!}\frac{\partial^n}{\partial z^n}f_j(x,z)\right|_{z=0}.
\end{equation}

\subsubsection{First relation}

Equation (\ref{de}) can be rewritten as

\begin{equation}
f_{j,n}(x)=\frac{1}{n!}\frac{\partial^{n-1}}{\partial
z^{n-1}}\left[\left(\prod_{k=1}^{g-1}\left(1+z~x^{m_k}\right)\right)\left(1+z~x^{m_g}\right)\right]\Big|_{z=0},
\end{equation}

\noindent and using Leibniz formula for the multiple derivative of a product of two functions, one obtains

\begin{equation}\label{int}
f_{j,n}(x)=\frac{1}{n!}\sum_{k=0}^{n}\bin{n}{k}\left[\frac{\partial^{n-k}}{\partial z^{n-k}}\prod_{p=1}^{g-1}\left(1+z~x^{m_p}\right)\right]\frac{\partial^{k}}{\partial z^{k}}\left(1+z~x^{m_g}\right)\Big|_{z=0},
\end{equation}

\noindent where $\bin{n}{k}=n!/k!/(n-k)!$ is the binomial coefficient. Equation (\ref{int}) yields

\begin{equation}
f_{j,n}(x)=\frac{1}{n!}\left[\frac{\partial^{n}}{\partial z^{n}}\prod_{k=1}^{g-1}\left(1+z~x^{m_k}\right)+x^{m_g}\frac{\partial^{n-1}}{\partial z^{n-1}}\prod_{k=1}^{g-1}\left(1+z~x^{m_k}\right)\right]\Big|_{z=0}.
\end{equation}

The number of states having angular momentum $J$ is given by relation (\ref{rec}) and $D\left(M,j,n\right)$ is the coefficient of $x^{M}$ in

\begin{equation}\label{summ}
f_{j,n}(x)=\sum_{M=M_{\mathrm{min}}}^{M_{\mathrm{max}}}D\left(M,j,n\right)~x^{M},
\end{equation}

\noindent which yields

\begin{equation}\label{recint}
D_g(M,j,n)=D_{g-1}(M,j,n)+D_{g-1}(M-m_g,j,n-1),
\end{equation}

\noindent where $D_g(M,j,n)$ represents the number of states with $n$ fermions (protons or neutrons) in $g$ one-fermion states. In a more general way, one can write the recursion relation (\ref{recint}) as 

\begin{equation}\label{mainresu0}
\left\{
\begin{array}{l}
D_k(M,j,n)=D_{k-1}(M,j,n)+D_{k-1}(M-m_k,j,n-1)\nonumber\\
D_k(0,j,n)=\delta(M)\;\;\;\forall k.
\end{array}
\right.
\end{equation}

\subsubsection{Second relation}

\noindent After a first derivation of $f_j(x,z)$, one gets

\begin{equation}
f_{j,n}(x)=\frac{1}{n!}\sum_{i=1}^gx^{m_i}\frac{\partial^{n-1}}{\partial
z^{n-1}}\prod_{k=1,k\ne i}^g\left(1+z~x^{m_k}\right)\Big|_{z=0},
\end{equation}

\noindent which is equivalent to

\begin{equation}
f_{j,n}(x)=\frac{1}{n!}\sum_{i=1}^gx^{m_i}\frac{\partial^{n-1}}{\partial z^{n-1}}\frac{\prod_{k=1}^g\left(1+z~x^{m_k}\right)}{\left(1+z~x^{m_i}\right)}\Big|_{z=0}.
\end{equation}

\noindent Using Leibniz formula, one obtains

\begin{eqnarray}
f_{j,n}(x)&=&\frac{1}{n!}\sum_{i=1}^gx^{m_i}\sum_{k=0}^{n-1}\bin{n-1}{k}\left[\frac{\partial^k}{\partial z^k}\frac{1}{\left(1+z~x^{m_i}\right)}\right]\frac{\partial^{n-1-k}}{\partial z^{n-1-k}}\prod_{p=1}^g\left(1+z~x^{m_p}\right)\Big|_{z=0}.
\end{eqnarray}

\noindent Since

\begin{equation}
\frac{\partial^k}{\partial z^k}\frac{1}{(1+az)}=\frac{k!(-1)^ka^k}{(1+az)^{k+1}},
\end{equation}

\noindent one finds

\begin{equation}\label{recu}
f_{j,n}(x)=\frac{1}{n}\sum_{k=1}^n(-1)^{k-1}\left[\sum_{i=1}^gx^{k.m_i}\right]f_{j,n-k}(x),
\end{equation}

\noindent which yields, in virtue of Eq. (\ref{summ}):

\begin{equation}\label{mainresu}
D\left(M,j,n\right)=\frac{1}{n}\sum_{k=1}^n\sum_{i=1}^g(-1)^{k-1}D\left(M-km_i,j,n-k\right).
\end{equation}

\noindent Such a formalism can be extended to include additional constraints \cite{PRATT00,PAIN11}. The number of loops required for the three-nested recursion relations (\ref{recint}) and (\ref{mainresu}) is roughly $n(2j+1)(2M_{\mathrm{max}}+1)$, \emph{i.e.} (number of fermions)$\times$(number of states)$\times$(number of values of $M$). The numerical cost is maximum for a half-filled shell, but the recursion relations are much more efficient than the usual combinatorial aproach since their cost is polynomial with $j$ and $n$.

\section{\label{sec3} Statistical modeling of $D(M,j,n)$}

The distribution $D(M,j,n)$ having a bell shape, it can be modeled as

\begin{equation}\label{stat}
D(M,j,n)=\frac{G\left(j^n\right)}{\sqrt{2\pi v\left(j^n\right)}}\exp\left[-\frac{M^2}{2v\left(j^n\right)}\right],
\end{equation}

\noindent where $G\left(j^n\right)=\bin{2j+1}{n}$ represents the degeneracy of $j^n$, and $v\left(j^n\right)$ its variance:

\begin{equation}\label{sig2}
v\left(j^n\right)=\sum_{M=-J_{\mathrm{max}}}^{J_{\mathrm{max}}}M^2=\frac{n(2j+1-n)(j+1)}{6}.
\end{equation}

\begin{figure}
\begin{center}
\includegraphics[width=8.5cm]{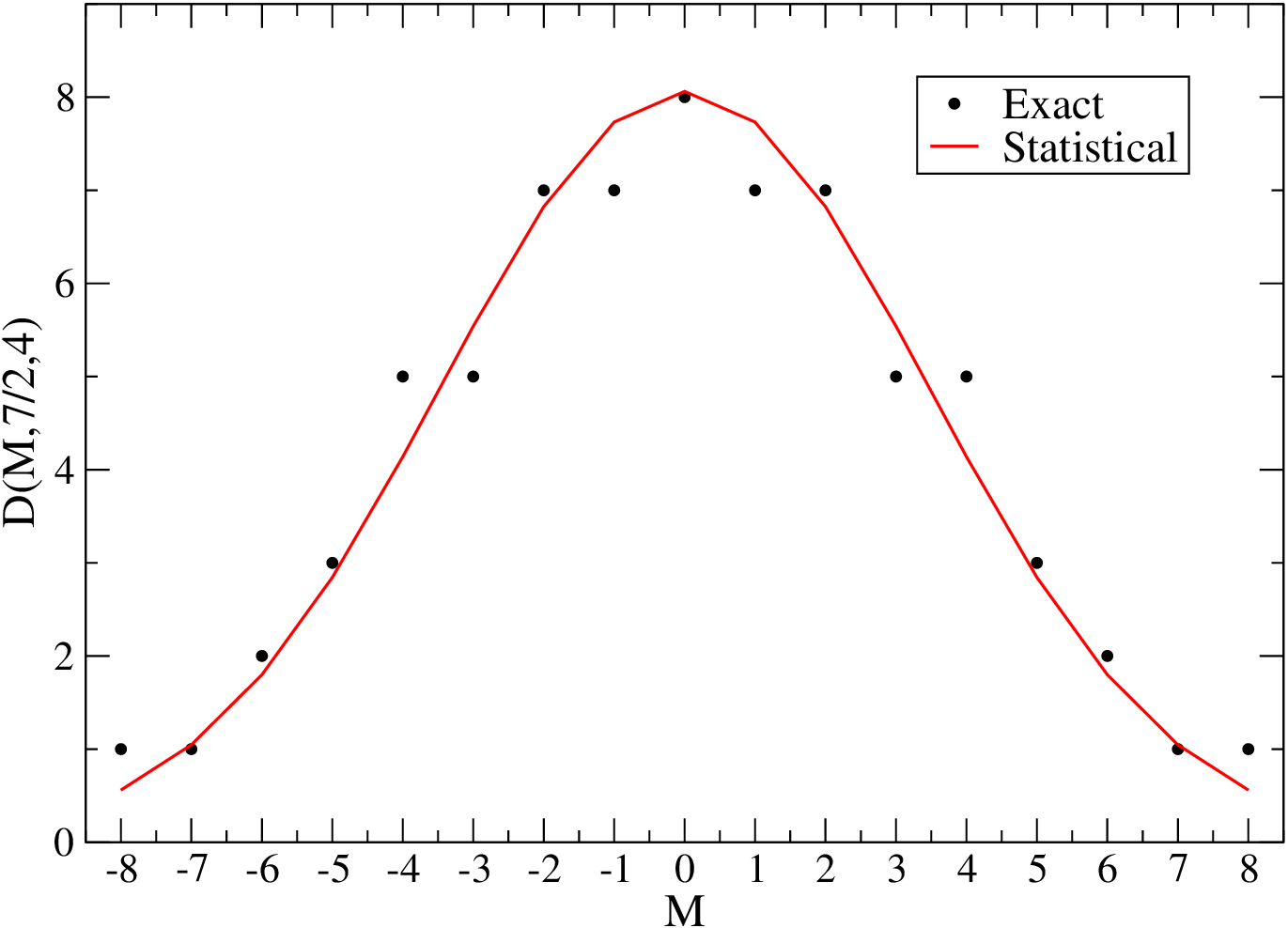}
\end{center}
\caption{Distribution $D(M,j,n)$ for $j$=7/2 and $n$=4: exact calculation (relation (\ref{mainresu})) and statistical modeling (Eq. (\ref{stat})).}
\label{3.5_4}
\end{figure}

\vspace{1cm}

\begin{figure}
\begin{center}
\includegraphics[width=8.5cm]{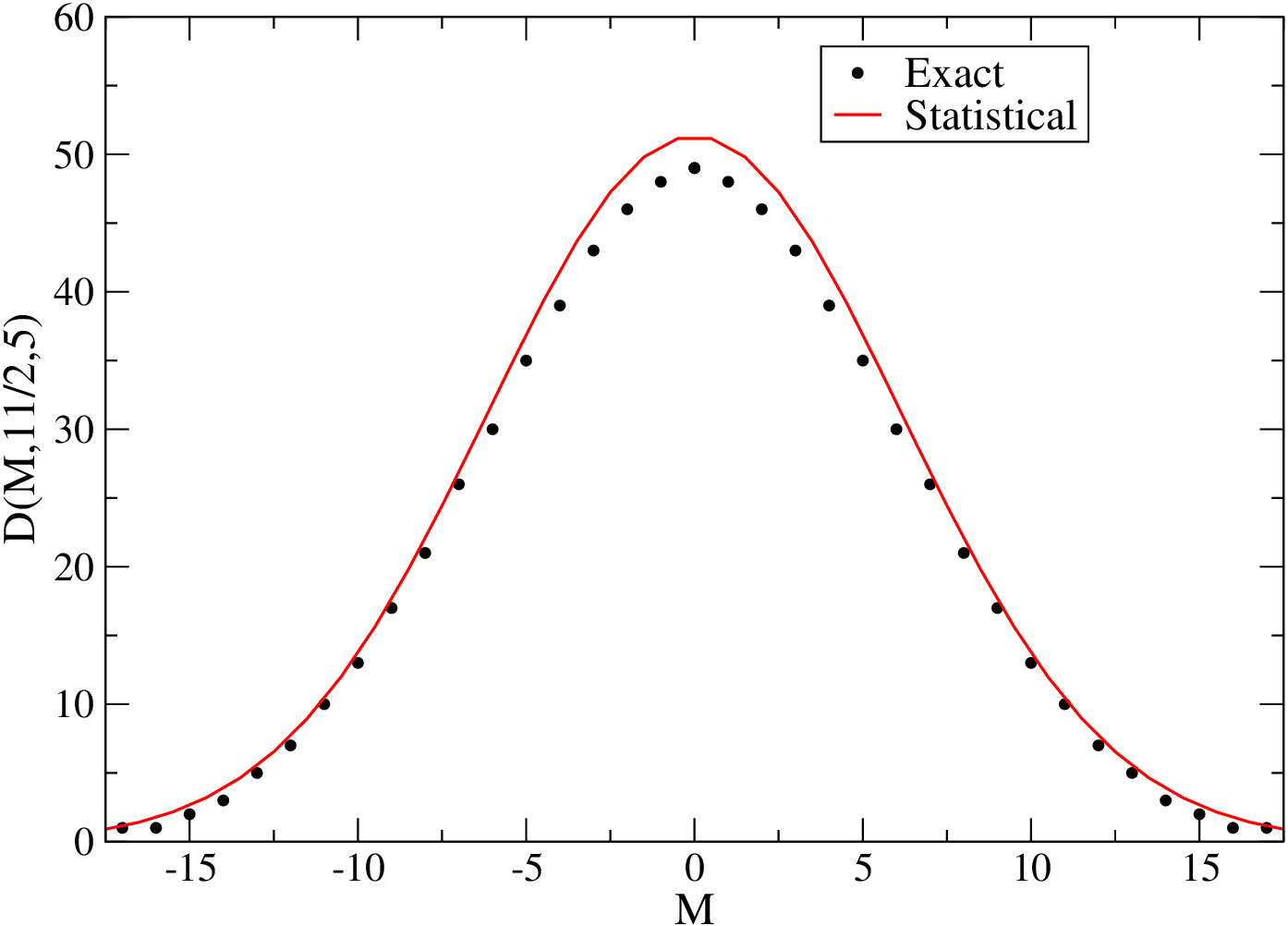}
\end{center}
\caption{Distribution $D(M,j,n)$ for $j$=11/2 and $n$=5: exact calculation (relation (\ref{mainresu})) and statistical modeling (Eq. (\ref{stat})).}
\label{5.5_5}
\end{figure}

\begin{figure}
\begin{center}
\vspace{5mm}
\includegraphics[width=8.5cm]{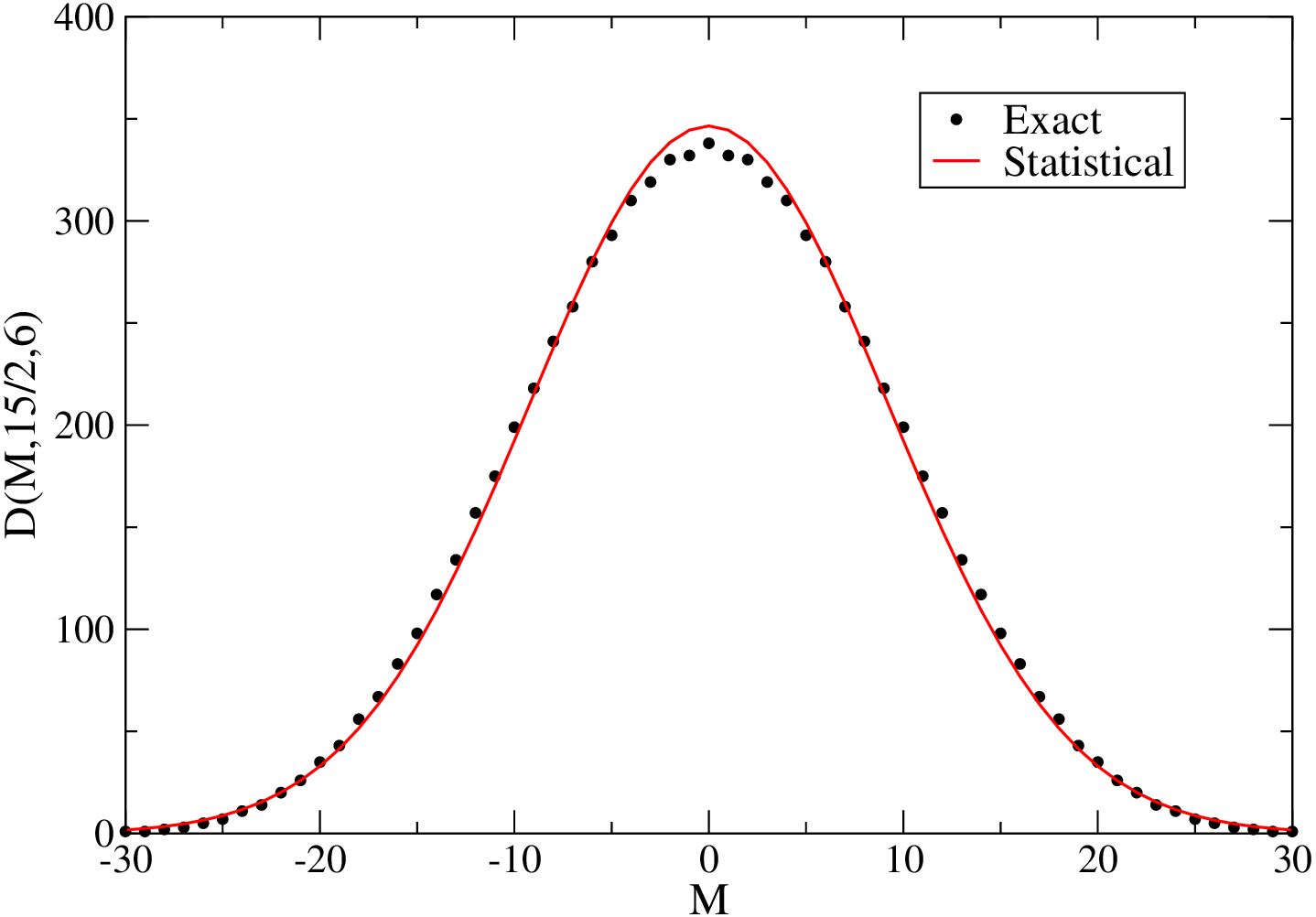}
\end{center}
\caption{Distribution $D(M,j,n)$ for $j$=15/2 and $n$=6: exact calculation (relation (\ref{mainresu})) and statistical modeling (Eq. (\ref{stat})).}
\label{7.5_6}
\end{figure}

\noindent One can see in figures \ref{3.5_4}, \ref{5.5_5} and \ref{7.5_6}, in the cases of shells $(7/2)^4$, $(11/2)^5$ and $(15/2)^6$ respectively, that the statistical modeling of $D(M,j,n)$ is in fairly good agreement with the exact distribution. The results can be improved taking into account the fourth-order moment (kurtosis), and a generalized Gaussian (or hyper-Gaussian) distribution. A first-order Taylor expansion of $J\rightarrow D(J,j,n)$ and $J\rightarrow D(J+1,j,n)$ at $J+1/2$ gives: 

\begin{equation}\label{plus12}
N(J,j,n)=D(J,j,n)-D(J+1,j,n)\approx\left.-\frac{dD}{dM}\right|_{M=J+1/2},
\end{equation}

\noindent and one gets, using Eq. (\ref{stat}):

\begin{equation}\label{qjs}
N(J,j,n)\approx\frac{G\left(j^n\right)}{\sqrt{2\pi}}\frac{(J+1/2)}{\left[v\left(j^n\right)\right]^{3/2}}\exp\left[-\frac{1}{2v\left(j^n\right)}\left(J+\frac{1}{2}\right)^2\right].
\end{equation}

\noindent It is interesting to compare the latter expression with the Ginocchio-Haxton formula

\begin{equation}\label{gh}
N(j,j,3)=\left[\frac{2j+3}{6}\right],
\end{equation}

\begin{figure}
\begin{center}
\includegraphics[width=8.5cm]{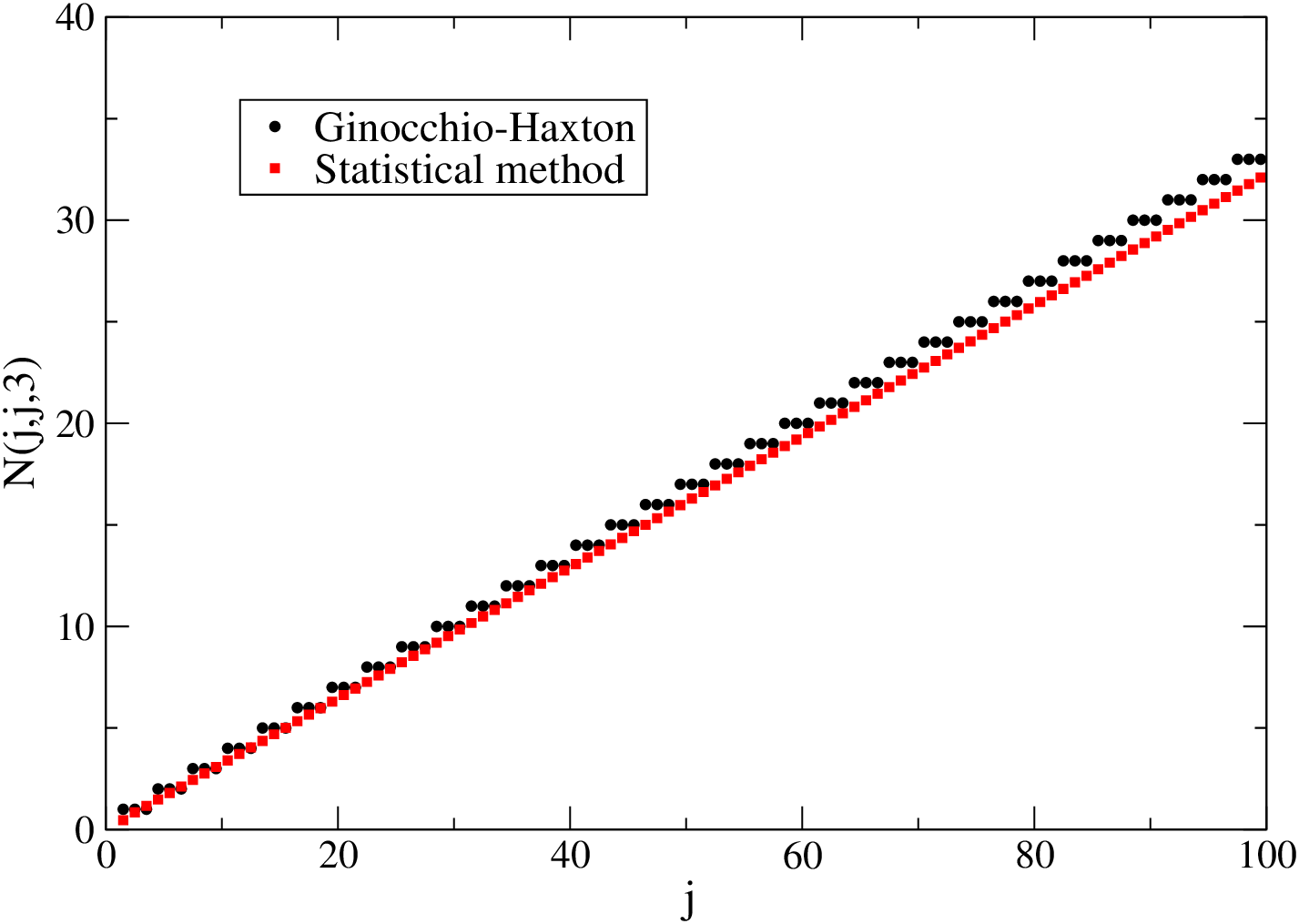}
\end{center}
\caption{Comparison between the exact Ginocchio-Haxton formula (relation (\ref{gh})) and statistical modeling (Eq. (\ref{qjs})) for $J=j$ and $n=3$.}
\label{haxton}
\end{figure}

\noindent where $[x]$ is the largest integer not exceeding $x$. One can see in Fig. \ref{haxton} that the results are rather close to the exact ones. The estimates can be improved either by performing the Taylor-series expansion up to a higher order in Eq. (\ref{plus12}), or using the expression:

\begin{equation}
N(J,j,n)=\int_{J-1/2}^{J+1/2}D(M,j,n)dM-\int_{J+1/2}^{J+3/2}D(M,j,n)dM.
\end{equation}

\noindent The statistical modeling is of course not as accurate as the recurrence relations, but it can be helpful to better understand the characteristics of the distribution of states and to derive, for instance, asymptotic expressions.

\section{\label{sec4} Excess of $J$ values}

\noindent Generating functions can also be of great interest for the determination of the excess of $J$ values, \emph{i.e.} the difference between the number of even values of $J$ and the number of odd values of $J$. For a configuration $j^n$ with $n=2k$, $k$ being a positive integer, the excess of $J$ values is equal to the excess of $M$ values. Since

\begin{equation}
f_{j,n}(x)=\sum_{M=M_{\mathrm{min}}}^{M_{\mathrm{max}}}D(M,j,n)~x^{M},
\end{equation}

\noindent the excess $E$ for a configuration $j^{n=2k}$ is equal to

\begin{eqnarray}
E\left(j^{2k}\right)&=&\sum_{M=M_{\mathrm{min}}}^{M_{\mathrm{max}}}(-1)^{M}D(M,j,n)=f_{j,n}(-1)=\left.\frac{1}{(2k)!}\frac{\partial^{2k}}{\partial z^{2k}}f_j(-1,z)\right|_{z=0}.
\end{eqnarray}

\noindent and the function $f_j(-1,z)$ is given by

\begin{equation}
f_j(-1,z)=\prod_{m=-j}^j\left[1+(-1)^mz\right]=\left(1+z^2\right)^{j+1/2},
\end{equation}

\noindent which implies

\begin{eqnarray}
E\left(j^{2k}\right)=\left.\frac{1}{(2k)!}\frac{\partial^{2k}}{\partial z^{2k}}\left[\sum_{p=0}^{j+1/2}\bin{j+1/2}{p}z^{2p}\right]\right|_{z=0}=\bin{j+1/2}{k}.
\end{eqnarray}

\noindent For two fermions (as well as for $n=2j+1-2=2j-1$), the number of odd-$J$ states is zero, since $J$ is necessarily even due to the Pauli exclusion principle (antisymmetric states). The values of the excess for different $j^n$ shells, relativistic or not, are displayed in table \ref{tab1}.

\begin{table}[ht]
\begin{center}
\begin{tabular}{cc}\hline\hline
$j^n$ & Excess \\\hline
$\left(1/2\right)^2$ & 1 \\ 
$\left(5/2\right)^4$ & 3 \\
$\left(7/2\right)^4$ & 6 \\
$\left(7/2\right)^6$ & 4 \\
$\left(9/2\right)^4$ & 10 \\
$\left(15/2\right)^6$ & 56 \\
\hline\hline
\end{tabular}
\end{center}
\caption{Excess of even-$J$ states for different $j^n$ shells.}\label{tab1}
\end{table}

\begin{table}[ht]
\begin{center}
\begin{tabular}{cccc}\hline\hline
Even $J$ & Number of states & Odd $J$ & Number of states \\\hline
$0$ & 1 & $1$ & 0 \\ 
$2$ & 2 & $3$ & 0\\
$4$ & 2 & $5$ & 1\\
$6$ & 1 & $7$ & 0\\
$8$ & 1 &     &   \\\hline
Total (even): & 7 & Total (odd): & 1\\\hline\hline
\end{tabular}
\end{center}
\caption{Number of even- and odd-$J$ states for the shell $(j=7/2)^4$.}\label{tab2}
\end{table}

\noindent In the case of $(7/2)^4$, there are 7 even-$J$ states and one odd-$J$ state, corresponding to $J$=5 (see table \ref{tab2}). The results can be checked with the tables published by Bayman and Lande \cite{BAYMAN66}. The numbers of even- and odd-$J$ states for the shell $(j=11/2)^n$ for different values of the number of fermions ($n$=4 and 6) are provided in table \ref{tab3}, and the number of states for all values of $J$ in tables \ref{tab4} (for $n=4$) and \ref{tab5} (for $n$=6). There is of course only one state with spin $J_{\mathrm{max}}$ (the expression of $J_{\mathrm{max}}$ is provided in Eq.(\ref{jmax})). It is worth mentioning that Talmi derived a recursion relation (which we recover using the generating-function formalism in the Appendix) and found interesting peculiarities in the distributions of spin-$J$ states: for instance, the states with spins $J_{\mathrm{max}}-2$ and $J_{\mathrm{max}}-3$ are unique in a $j^n$ configuration and there is no state with spin $J_{\mathrm{max}}-1$ \cite{TALMI05}.
           
\begin{table}[ht]
\begin{center}
\begin{tabular}{cccc}\hline\hline
$n$\;\;\;\; & Number of \;\;\;\; & Number of \;\;\;\; & Excess \;\;\;\; \\
 \;\;\;\;   & even-$J$ states \;\;\;\; & odd-$J$ states \;\;\;\; & \\\hline
$4$ \;\;\;\; & 24 \;\;\;\; & 9 \;\;\;\; & 15 \;\;\;\;\\ 
$6$ \;\;\;\; & 39 \;\;\;\; & 19 \;\;\;\; & 20 \;\;\;\;\\
\hline\hline
\end{tabular}
\end{center}
\caption{Number of even- and odd-$J$ states for the shell $(j=11/2)^n$ for $n$=4 and 6.}\label{tab3}
\end{table}
     
\begin{table}[ht]
\begin{center}
\begin{tabular}{cccc}\hline\hline
Even $J$ & Number of states & Odd $J$ & Number of states\\\hline
$0$ & 2 & $1$ & 0\\ 
$2$ & 3 & $3$ & 1\\
$4$ & 4 & $5$ & 2\\
$6$ & 4 & $7$ & 2\\
$8$ & 4 & $9$ & 2\\
$10$ & 3 & $11$ & 1\\
$12$ & 2 & $13$ & 1\\
$14$ & 1 & $15$ & 0\\
$16$ & 1 &      &\\\hline
Total (even): & 24 & Total (odd): & 9\\\hline\hline
\end{tabular}
\end{center}
\caption{Number of even- and odd-$J$ states for the shell $(j=11/2)^4$. The excess is equal to 15.}\label{tab4}
\end{table}

\begin{table}[ht]
\begin{center}
\begin{tabular}{cccc}\hline\hline
Even $J$ & Number of states & Odd $J$ & Number of states\\\hline
$0$ & 3 & $1$ & 0\\ 
$2$ & 4 & $3$ & 3\\
$4$ & 6 & $5$ & 3\\
$6$ & 7 & $7$ & 4\\
$8$ & 6 & $9$ & 4\\
$10$ & 5 & $11$ & 2\\
$12$ & 4 & $13$ & 2\\
$14$ & 2 & $15$ & 1\\
$16$ & 1 & $17$ & 0\\
$18$ & 1 &      &\\\hline
Total (even): & 39 & Total (odd): & 19\\\hline\hline
\end{tabular}
\end{center}
\caption{Number of even- and odd-$J$ states for the shell $(j=11/2)^6$. The excess is equal to 20.}\label{tab5}
\end{table}

\noindent It is interesting to evaluate, for $2<n<2j-$1, the ratio between the excess $E$ and the total number of fixed-spin states $N_{\mathrm{tot}}$ for specific configuration $j^{2k}$:

\begin{equation}
r\left(j^{2k}\right)=\frac{E\left(j^{2k}\right)}{N_{\mathrm{tot}}\left(j^{2k}\right)},
\end{equation}

\noindent where $N_{\mathrm{tot}}$ reads

\begin{equation}
N_{\mathrm{tot}}=\sum_{J=J_{\mathrm{min}}}^{J_{\mathrm{max}}}N(J,j,n).
\end{equation}

\noindent The latter quantity can be approximated by 

\begin{equation}
N_{\mathrm{tot}}\approx \frac{G\left(j^{2k}\right)}{\sqrt{2\pi v\left(j^{2k}\right)}},
\end{equation}

\noindent where $G\left(j^{2k}\right)=\bin{2j+1}{2k}$ is the degeneracy of configuration $j^{2k}$ and $ v\left(j^{2k}\right)$ the variance (see Eq.(\ref{sig2})):

\begin{equation}
v\left(j^{2k}\right)=\frac{k(2j+1-2k)(j+1)}{3}.
\end{equation}

\noindent One has therefore

\begin{equation}
r\left(j^{2k}\right)\approx\frac{\bin{j+1/2}{k}}{\bin{2j+1}{2k}}\sqrt{\frac{2\pi k(2j+1-2k)(j+1)}{3}}.
\end{equation}

\begin{figure}
\begin{center}
\includegraphics[width=8.5cm]{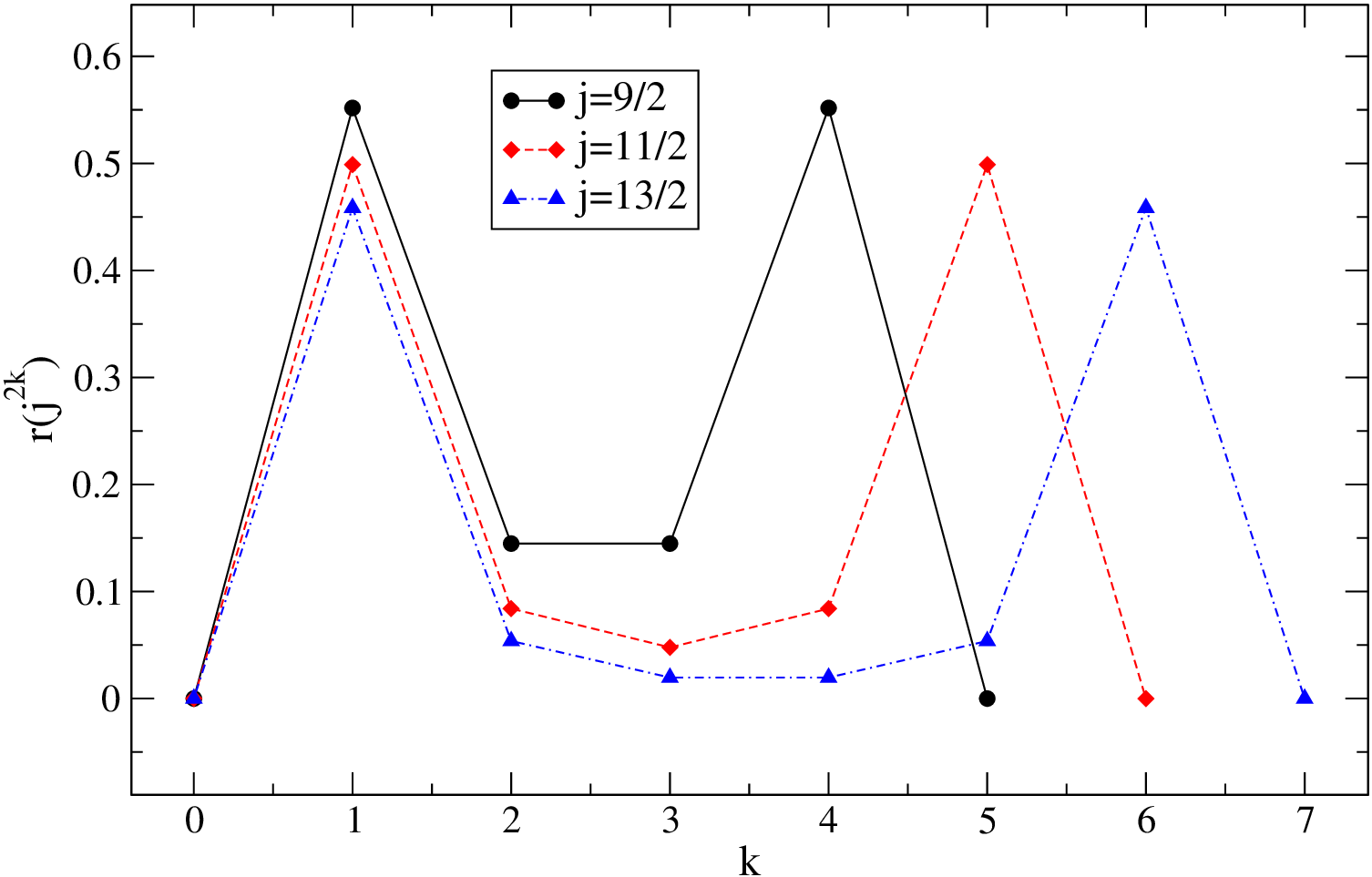}
\end{center}
\caption{Values of $r\left(j^{2k}\right)$ as a function of $k$ for $j$=9/2, 11/2 and 13/2.}
\label{figr}
\end{figure}

\noindent Such a quantity reaches its minimum 

\begin{equation}
r_{\mathrm{min}}=\frac{\sqrt{j(1+4j(j+2))-3}~\Gamma\left(\frac{j}{2}+\frac{1}{4}\right)\Gamma\left(\frac{j}{2}+\frac{5}{4}\right)}{2\sqrt{3}j!}
\end{equation}

\noindent for $k=j/2-1/4$ if $j$ is of the form $j=2p+1/2$ with $p$ a positive integer ($\Gamma$ is the usual Gamma function) and

\begin{equation}
r_{\mathrm{min}}=\frac{(2j+1)\sqrt{(j+1)}~\Gamma\left(\frac{j}{2}+\frac{3}{4}\right)^2}{2\sqrt{3}j!}
\end{equation}

\noindent for $k=j/2+1/4$ if $j$ is of the kind $j=2p+3/2$. Figure \ref{figr} represents the quantity $r\left(j^{2k}\right)$ for $j$=9/2, 11/2 and 13/2.

Using Stirling formula, one finds that the degeneracy of $j^n$ at the maximum complexity ($n=j+1/2$) varies as

\begin{equation}
G\left(j^{j+1/2}\right)\approx 2^{2j+1}
\end{equation}

\noindent and the following asymptotic form for $j\rightarrow\infty$ is obtained:

\begin{equation}
r_{\mathrm{min}}\approx\sqrt{\frac{2\pi}{3}}\frac{j^{3/2}}{2^j}.
\end{equation}

\section{Conclusion}

Using the generating-function formalism, new recursion relations were derived for the number of antisymmetric states with a given value of $J$ due to the coupling of $n$ identical fermions in the $j$ orbit. Still using the generating function, an odd-even staggering was found in the spin distribution of a single-j shell with an even number of fermions. The excess of the number of states with an even value of $J$ was calculated and its asymptotic behavior for large values of $j$ was investigated using a statistical modeling of the number of states with angular-momentum projection $M$. 

\appendix

\section{Talmi's recursion relation}

The generating-function formalism enables one to derive another recursion relation. Indeed, according to Eq. (\ref{eqfj}), one has

\begin{eqnarray}
f_{j+1}(x,z)&=&\prod_{m=-j-1}^{j+1}\left(1+x^mz\right)=\sum_{n=0}^{2j+3}z^nf_{j+1,n}(x)\nonumber\\
&=&f_j(x,z)\left(1+z~x^{-j-1}\right)\left(1+z~x^{j+1}\right)\nonumber\\
& &
\end{eqnarray}

\noindent and therefore

\begin{eqnarray}
f_{j+1,n}(x)=\left.\frac{1}{n!}\frac{\partial^n}{\partial z^n}f_{j+1}(x,z)\right|_{z=0}=\frac{1}{n!}\sum_{k=0}^n\bin{n}{k}\frac{\partial^{n-k}}{\partial z^{n-k}}f_j(x,z)\left.\frac{\partial^k}{\partial z^k}\mathcal{P}_j(x,z)\right|_{z=0},
\end{eqnarray}

\noindent where

\begin{eqnarray}\label{quad}
\mathcal{P}_j(x,z)&=&\left(1+z~x^{-j-1}\right)\left(1+z~x^{j+1}\right)=1+z^2+z\left(x^{-j-1}+x^{j+1}\right).
\end{eqnarray}

\noindent Since expression (\ref{quad}) is a second-order polynomial, the only derivatives which are non zero correspond to

\begin{equation}
\left\{\begin{array}{ll}
k=0: & \left.\frac{\partial^0}{\partial z^0}\mathcal{P}_j(x,z)\right|_{z=0}=1\\
k=1: & \left.\frac{\partial^1}{\partial z^1}\mathcal{P}_j(x,z)\right|_{z=0}=x^{-j-1}+x^{j+1}\\
k=2: & \left.\frac{\partial^2}{\partial z^2}\mathcal{P}_j(x,z)\right|_{z=0}=2,\\
\end{array}\right.
\end{equation}

\noindent which leads to

\begin{eqnarray}
D\left(M,j+1,n\right)&=&D\left(M,j,n\right)+D\left(M-j-1,j,n-1\right)\nonumber\\
& &+D\left(M+j+1,j,n-1\right)+D\left(M,j,n-2\right),
\end{eqnarray}

\noindent from which a recursion relation can be deduced for $N$:

\begin{eqnarray}
N\left(J,j+1,n\right)&=&N\left(J,j,n\right)+N\left(J-j-1,j,n-1\right)\nonumber\\
& &+N\left(J+j+1,j,n-1\right)+N\left(J,j,n-2\right),
\end{eqnarray}

\noindent \emph{i.e.}, in a compact form

\begin{eqnarray}
N\left(J,j+1,n\right)=\sum_{i,k=0}^1N\left(J+(i-k)(j+1),j,n-i-k\right),
\end{eqnarray}

\noindent which is the relation (5) derived by Talmi in Ref. \cite{TALMI05} for $J\geq j$.

\end{document}